\documentclass{aa}
\usepackage{graphics}

\newcommand{\etal}{{\em et al.}}

\newcommand{\mic}{\mbox{$\mu$m}}

\begin{document}

\thesaurus{11.01.2; 11.06.1; 11.09.1 Centaurus A; 11.09.2; 11.10.1; 11.19.2; 13.09.1; 13.18.1}

\title{A barred spiral at the centre of the giant elliptical radio 
galaxy 
Centaurus A\thanks{Based on observations with ISO, an ESA project with 
instruments funded by ESA Member States (especially the PI countries: 
France, 
Germany, the Netherlands and the United Kingdom) and with participation 
of 
ISAS and NASA.}}

\author{I.F. Mirabel\inst{1,2}, 
O. Laurent\inst{1}, 
D.B. Sanders\inst{3}, 
M. Sauvage\inst{1}, 
M. Tagger$^1$,  
V. Charmandaris\inst{4}, 
L. Vigroux\inst{1},
P. Gallais \inst{1},
C. Cesarsky \inst{1},  
\& D.L. Block\inst{5}} 

\institute{
CEA/DSM/DAPNIA Service d'Astrophysique F-91191 Gif-sur-Yvette, France
\and
IAFE. cc 67, suc 28. 1428 Buenos Aires. Argentina
\and 
Institute for Astronomy, University of Hawaii, 2680 Woodlawn Drive, 
Honolulu, HI 96822
\and
Observatoire de Paris, DEMIRM, 61 Av. de l'Observatoire, F-75014 Paris, 
France
\and
Department of Computational and Applied Mathematics, University of the 
Witwatersrand, Private Bag 3, WITS 2050, South Africa
}

\titlerunning{Spiral structure at the centre of Centaurus A}

\authorrunning{Mirabel \etal}

\date{Received October 7, 1998/ Accepted October 21, 1998}

\offprints{I.F. Mirabel, mirabel@discovery.saclay.cea.fr} 

\maketitle

\begin{abstract}

We report observations at mid-infrared and sub-millimeter wavelengths
of Centaurus A (CenA, NGC 5128), the giant elliptical galaxy that
harbors the closest radio loud Active Galactic Nucleus (AGN) to Earth.
The dust emission from the deep interior of CenA reveals a bisymmetric
structure with a diameter of 5$^{\prime}$ (5 kpc), centred at the
AGN. This structure is remarkably similar to that of a barred spiral
galaxy, with the bar lying in a plane that is tilted $\sim$
18$^{\circ}$ from the line of sight. The true nature of the
distribution of dust in the inner regions of CenA is noticeably
displaced from the more chaotic and widespread optical
obscuration. The barred spiral is a quasi-stable structure formed at
the center of the giant elliptical from the tidal debris of a gas-rich
object(s) accreted in the past 10$^9$ years. The total size and mass
of interstellar gas in the barred spiral at the center of CenA is
comparable to the small Local Group spiral galaxy Messier 33. The
observation of this remarkable structure opens the more general
question on whether the dusty hosts of giant radio galaxies like CenA,
are ``symbiotic" galaxies composed of a barred spiral inside an
elliptical, where the bar serves to funnel gas toward the AGN.

\keywords{Galaxies: active -- Galaxies: formation -- Galaxies: individual: NGC 5128 (Centaurus A) -- Galaxies: interactions -- Galaxies: jets -- Galaxies: spiral -- Infrared: galaxies -- Radio continuum: galaxies} 

\end{abstract}


\section{Introduction}

Radio galaxies are thought to be giant ellipticals powered by
accretion of interstellar matter onto a supermassive black
hole. Interactions with gas rich galaxies may feed them with cold
interstellar matter, but a major problem is how the fueling gas finds
its way from typical galactic radii of several kiloparsecs down to a
few parsecs from the centre, and at the rates required to power the
AGN. Bar structures have been proposed in theoretical models
(\cite{Shlosman}) as the dynamical instabilities that deliver fuel to
the AGN. However, the observation of bars of cold gas and dust in the
deep interior of luminous giant ellipsoidal systems of stars that
hosts a powerful radio source has been difficult.

CenA may serve as a template to investigate these questions. The
prominent dark bands seen across it suggest that it is the product of
the merger between a small gas-rich galaxy with a larger elliptical
(\cite{Baade}). The radio emission comes primary from two large lobes
separated by 5$^{\circ}$ on the sky ($\sim$ 300 kpc at a distance of
3.5 Mpc). The lobes are powered by relativistic jets that emanate from
a central region $\sim$ 0.01 pc in size, widely believed to contain a
massive black hole (\cite{Kellerman}). The center is hidden behind
large columns of gas and dust with visual extinctions that reach
values as large as A$_V$ = 70 mag. The absorption is less critical in
the near infrared bands (1-2$\mu$m), but at these wavelengths the
contribution to the total flux from old stars belonging to the
ellipsoidal galaxy is a major, if not dominant contributor to the
total flux.  It is in the mid-infrared and longer wavelengths that the
emission from dust in the deep interior of a giant ellipsoidal system
can be better traced.

\vfill
\section{ISO and SCUBA observations}

Mid-infrared observations were made with the Infrared Space
Observatory (\cite{Kessler}) Camera ISOCAM (\cite{Cesarsky}) using the
broad-band filters LW2 (5.0-8.5 $\mu$m, $\lambda$$_0$ = 6.75 $\mu$m)
and LW3 (12-18 $\mu$m, $\lambda$$_0$ = 15 $\mu$m), and a Circular
Variable Filter (CVF) covering the range 5-16 $\mu$m. The LW2 and LW3
observations were made with a 3\arcsec/pixel lens resulting in full
widths at half-maximum (FWHM) of 4\arcsec and 6\arcsec respectively.
The CVF observations were done with a 1.5\arcsec/pixel lens resulting
in full widths at half-maximum (FWHM) in the range of 3-6\arcsec \ts
for wavelengths of 5-18 $\mu$m.  The standard data reduction
procedures described in the ISOCAM\footnote{The ISOCAM data presented
in this paper were analyzed using ``CIA'', a joint development by the
ESA Astrophysics Division and the ISOCAM Consortium led by the ISOCAM
PI, C. Cesarsky, Direction de Sciences de la Mati\`ere, C.E.A.,
France.}  manual were followed (\cite{isoman}). Dark subtraction was
performed using a model of the secular evolution of ISOCAM's dark
current (\cite{biviano}). Cosmic rays were removed using a
multi-resolution median filtering method (\cite{starck}) while the
memory effects of the detector were corrected using the so-called IAS
transient correction algorithm which is based on an inversion method
(\cite{abergel}). The final raster was constructed after using the
instrumental flat fields and correcting for the lens field
distortion. These methods and their consequences are discussed in
detail in Starck et al. (1998).

Submillimeter observations at 450 $\mu$m and 850 $\mu$m were obtained
using the Submillimeter Common-User Bolometer Array (SCUBA,
\cite{Cunningham}) at the James Clerk Maxwell Telescope (JCMT) on
Mauna Kea.  Observations at both wavelengths were obtained
simultaneously using the 91 element shortwavelength array at 450
$\mu$m (HPBW = 9.5\arcsec) and the 37 element longwavelength array at
850 $\mu$m (HPBW = 14.7\arcsec).

\section{The dust emission}

\begin{figure*}[h]
\resizebox{18cm}{!}{\includegraphics{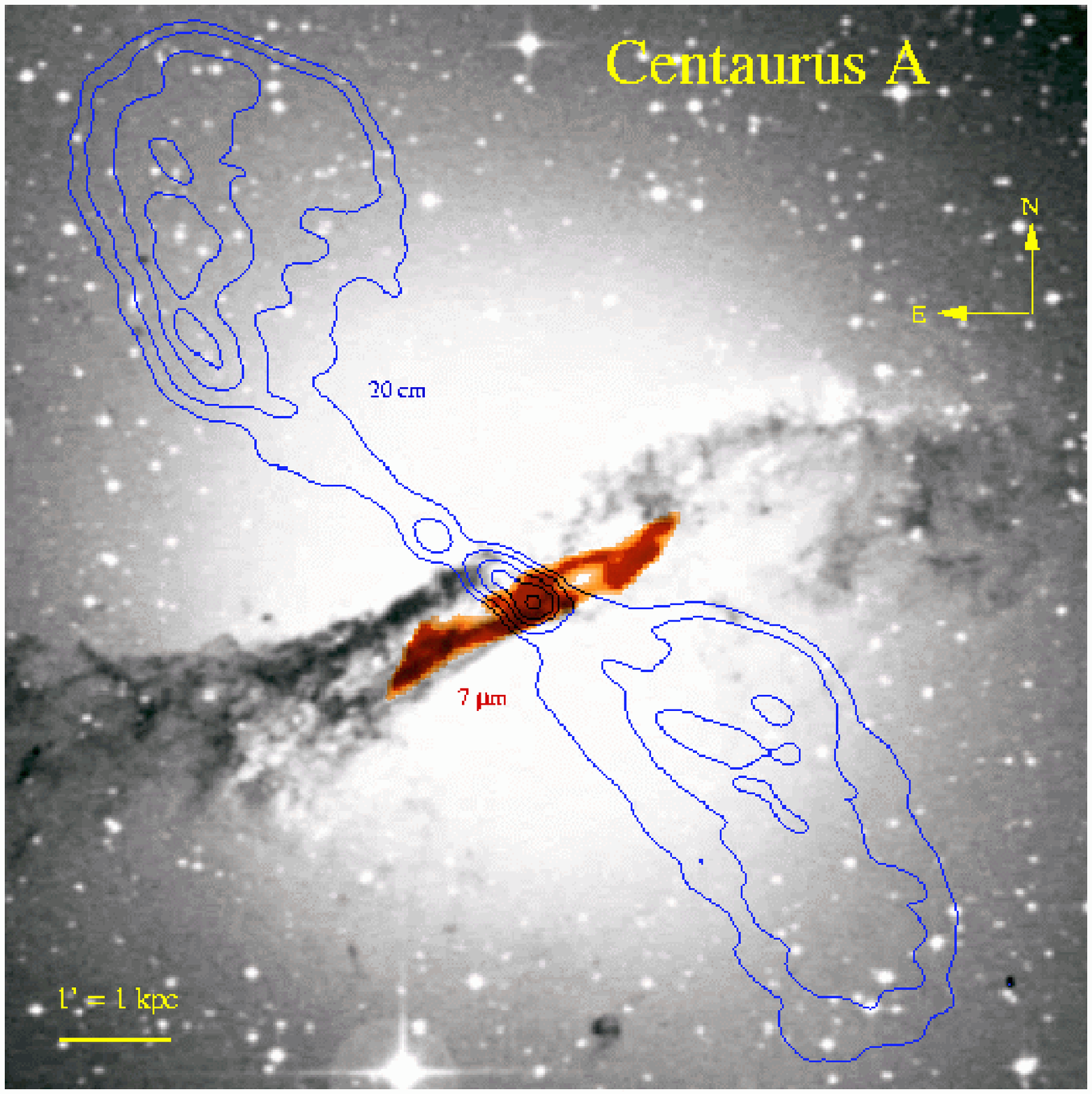}}
\caption{
The ISO 7{\ts}$\mu$m emission (red image) and VLA 20 cm continuum
(\cite{Condon}) (blue contours), overlaid on an optical image (black
and white) from the Palomar Digital Sky Survey. The emission from dust
with a bisymmetric morphology at the centre is about 10 times smaller
than the overall size of the shell structure in the elliptical
(\cite{Malin}) and lies on a plane that is almost parallel to the
minor axis of its giant host. Whereas the gas associated to the spiral
rotates with a maximum radial velocity of 250 km s$^{-1}$, the
ellipsoidal stellar component rotates slowly approximately
perpendicular to the dust lane (\cite{Wilkinson}). The synchrotron
radio jets shown in this figure correspond to the inner structure of a
double lobe radio source that extends up to 5$^{\circ}$ ($\sim$ 300
kpc) on the sky. The jets are believed to be powered by a massive
black hole located at the common dynamic center of the elliptical and
spiral structures.
}
\label{iso_dss}
\end{figure*}

Figure \ref{iso_dss} shows the mid-infrared, radio, and optical images
of CenA.  The 7 $\mu$m emission from dust reveals a bisymmetric
structure of 5$^\prime$ ($\sim${\ts}5 kpc for a distance of 3.5 Mpc)
in total length. In contrast to the optical dark lanes which show a
wide and somewhat chaotic distribution, the structure of the
mid-infrared emission is remarkably thin, smooth and bisymmetric.

\begin{figure*}
\resizebox{\hsize}{!}{\includegraphics{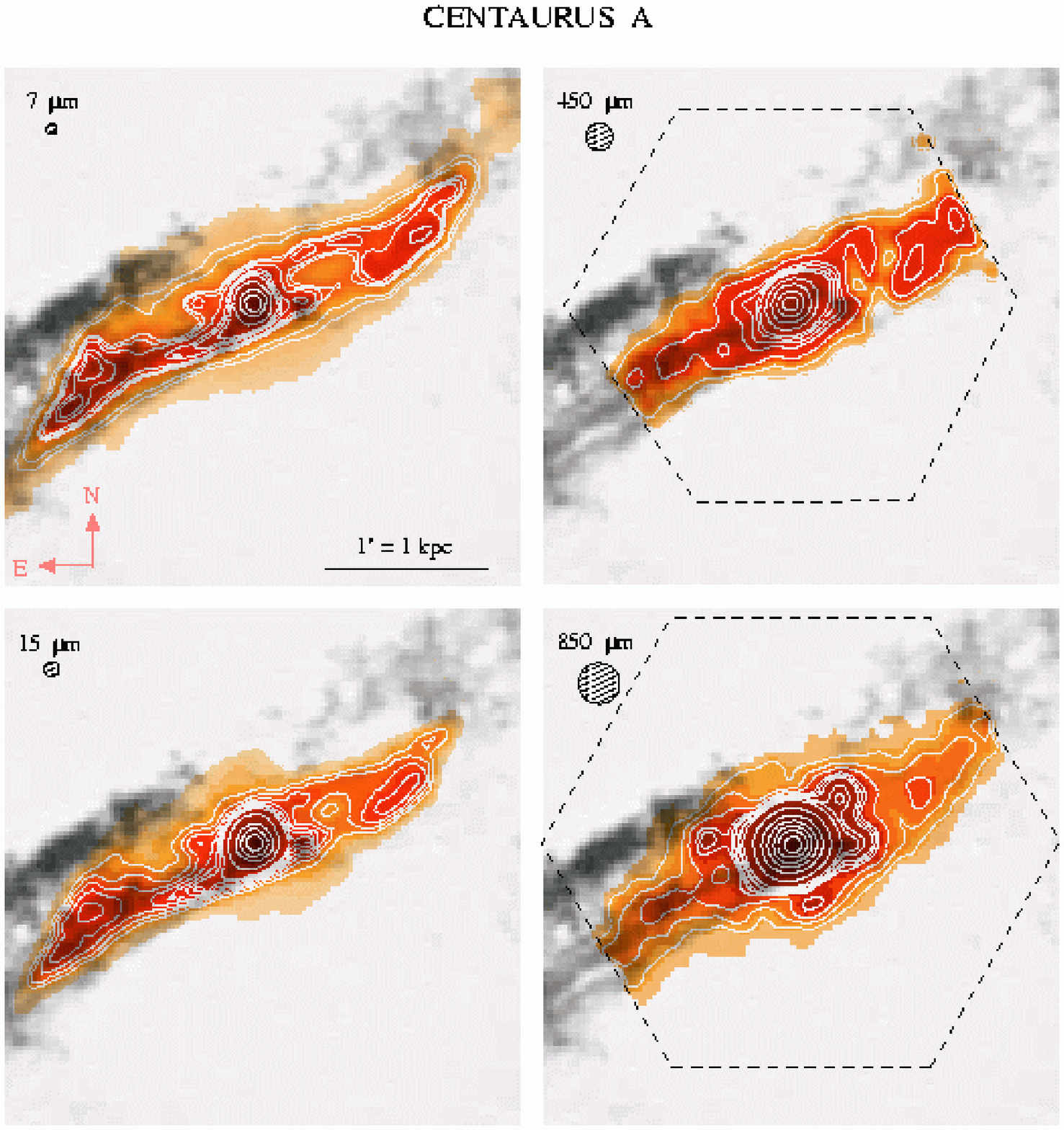}}
\caption{
Distribution of dust in the inner region of CenA as observed at
mid-infrared and submillimeter wavelengths, with optical dark lanes
superimposed. In the outer regions the 7{\ts}$\mu$m and 15{\ts}$\mu$m
disks exhibit anticlockwise twists with increasing radius, whereas the
dark lanes in the optical image twist clockwise. No submillimeter
emission was detected from the most prominent dark lanes. For the
observations with ISOCAM were used the broad-band filters LW2
(5.0--8.5{\ts}$\mu$m, $\lambda$$_0$ = 6.75{\ts}$\mu$m) and LW3
(12--18{\ts}$\mu$m, $\lambda$$_0$ = 15{\ts}$\mu$m). The LW2 and LW3
observations were made with a 3\arcsec/pixel lens resulting in full
widths at half-maximum (FWHM) of 4\arcsec\ and 6\arcsec\ respectively.
The observations at 450{\ts}$\mu$m and 850{\ts}$\mu$m were obtained
simultaneously using the 91 element shortwavelength array at
450{\ts}$\mu$m (HPBW = 9.5\arcsec) and the 37 element longwavelength
array at 850{\ts}$\mu$m (HPBW = 14.7\arcsec). The hexagons in the
third and fourth panels indicate the limits of the SCUBA maps.
Contours for the 7{\ts}$\mu$m map are: 2 2.5 3 3.5 4 4.5 5 5.5 7 11 18
24 mJy arcsec$^{-2}$, for the 15{\ts}$\mu$m map: 2 2.5 3 3.5 4 4.5 5
5.5 7 11 18 29 mJy arcsec$^{-2}$, for the 450{\ts}$\mu$m map: 41 58 83
116 133 150 166 232 331 464 580 mJy arcsec$^{-2}$, for the
850{\ts}$\mu$m map: 10 14 18 22 24 25 29 30 36 72 145 218 290 363 436
mJy arcsec$^{-2}$.
}
\label{iso_scuba}
\end{figure*}

Figure \ref{iso_scuba} shows maps of CenA in the mid-infrared and
submillimeter wavelengths. The emission at 7 $\mu$m and 15 $\mu$m
observed with ISO comes from small dust grains (radii 0.05 $\mu$m and
less) that can undergo large temperature (T$_d$ $\geq$ 100 K)
excursions (\cite{Desert}). On the contrary, the 450 $\mu$m and 850
$\mu$m emission is mostly due to the large and cold (T$_d$ $\sim$
10-20 K) dust grains which dominate the extinction of visible
light. The warm dust is clearly displaced from the most prominent
optical dark lanes; in the outer regions the 7 $\mu$m and 15 $\mu$m
disks exhibit anticlockwise twists with increasing radius, whereas the
dark lanes in the optical image twist clockwise.  The emitting dust is
less extended and clearly displaced from the most prominent optical
dark lanes.

Figure \ref{iso_scuba} shows that these two dust components have the
same general distribution, and therefore the displacement of the
emitting dust from the optical dark lanes in Figures \ref{iso_dss} and
\ref{iso_scuba} is not due to major differences between the spatial
distributions of the cold and very warm dust components.  In a three
dimensional tilted and warped disk, projection effects play an
important role. The optical appearance of the dark lanes in a luminous
ellipsoidal system may be strongly affected by relatively small
amounts of cold dust in the outer parts of the bending disk located in
the foreground side of the luminous ellipsoidal distribution of stars.

Table 1 gives the infrared/submillimeter fluxes for the inner nuclear
region (radius $\leq${\ts}7.5\arcsec) and the bar (annulus of
7.5\arcsec--60\arcsec). The infrared and submillimeter fluxes from the
nuclear region are typically less than 10\% of the total flux from the
annulus of 7.5\arcsec-60\arcsec {\ts} radius mapped with
SCUBA. Assuming $M_d=\frac{f_{\nu} D^2}{\kappa_{\nu} B_{\nu}(T_d)}$
where $\kappa_{\nu}=40(\nu/3\ts 10^{12}$Hz)$^n$ with n=1.5-2 and D the
distance, the combination of the ISOCAM, SCUBA and IRAS 60 $\mu$m and
100 $\mu$m fluxes defines a spectral energy distribution for the
annulus that can be fitted by two dust components: T$_d$ = 35 K and
M$_d$ = 7 10$^4$ M$_{\odot}$ and T$_d$ = 12 K and M$_d$ = 4 10$^7$
M$_{\odot}$. Such large masses of cold dust are observed in spiral
galaxies (\cite{Alton, Krugel}), once the spectrum at $\geq$ 100
$\mu$m is taken into account.

The far-infrared luminosity of CenA is 8 10$^9$ L$_{\odot}$, and when
combined with a total H$_2$ mass of 3.4 10$^8$ M$_{\odot}$ results in
a L$_{FIR}$/M(H$_2)$ of 24 L$_{\odot}$/M$_{\odot}$ (\cite{Eckart2}),
which is typical of starburst galaxies (\cite{Sanders}).

\begin{figure}
\resizebox{\hsize}{!}{\includegraphics{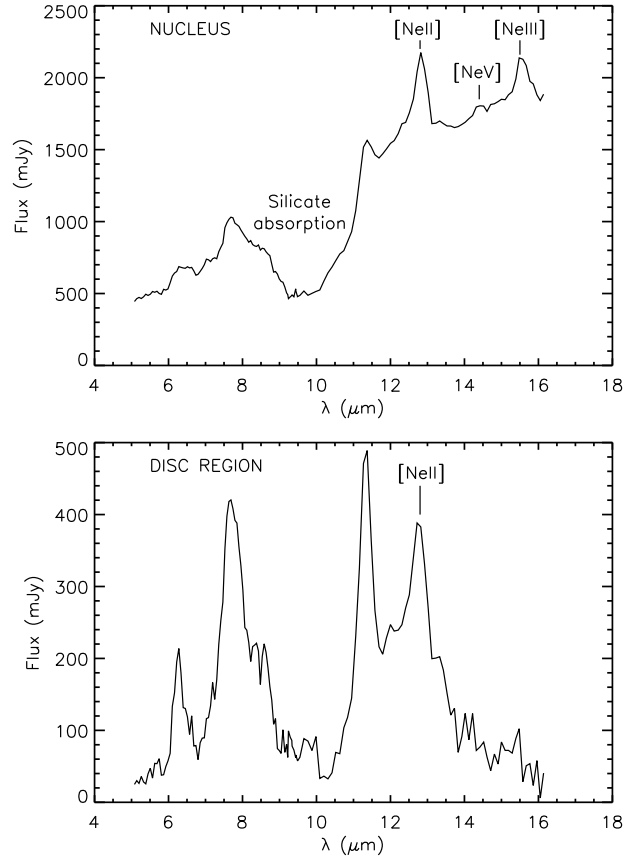}}
\caption{
ISOCAM CVF spectra for a region of 4\arcsec\ radius around the bright
unresolved infrared nucleus (upper panel), and for a typical starburst
spot in the primary bar (lower panel). The spectrum of the nucleus is
typical of mid-infrared spectra of AGNs which always show a strong
continuum flux below 6{\ts}$\mu$m associated with faint or absent PAH
emission (\cite{Lutz}). The [Ne III]/[Ne II] emission line ratio which
is a measure of the UV hardness if the radiation field is harder in
the nucleus than anywhere else in CenA. On the contrary, the
extra-nuclear region presents strong PAH bands and a faint continuum
in agreement with a lower radiation field.
}
\label{spectra}
\end{figure}

The ISOCAM broad band spectrum of the nucleus, shown in Figure
\ref{spectra}, is typical of the mid-infrared spectra of AGNs, whereas
the spectra of the bright spots in the arcs and arms are typical of
photo-dissociation regions in spiral galaxies.  The main distinct
features in the mid-infrared spectrum of the nucleus relative to the
bright spots in the disk are: 1) a noticeable continuum flux below
6{\ts}$\mu$m that seems to be present in every AGN, 2) a strong and
steeply rising continuum as a function of wavelength associated with a
faint PAH emission perhaps due to contamination by the bar, 3) deep
absorption from silicates around 10{\ts}$\mu$m, 4) [Ne III] and [Ne V]
emission lines detected in the nuclear region only. The [Ne V]
high-excitation line which is predominantly powered by hard AGN
radiation fields (\cite{Genzel}) is only present in the nuclear region
of CenA.

\begin{table}[h]
\caption[]{Mid-Infrared (ISO) and Submillimeter (SCUBA) Fluxes}
\label{tab:data}
\begin{tabular}{lcccc}
  \hline
   Filter & $\lambda_{\rm m}$ & $f_{\rm nucleus}$ & $f_{\rm disk}$ & 
$f_{\rm 
total}$ \\ 
          &         ($\mu$m)  &          (Jy)  &          (Jy)  &            
(Jy) \\
  \hline
  \hline
   ISOCAM LW2   & 7   &  0.6 &  8.8  &   9.4 \\
   IRAS~12$^a$  & 12  &  --- &  ---  &  11.2 \\
   ISOCAM LW3   & 15  &  1.2 & 10.8  &  12.0 \\
   IRAS~25      & 25  &  --- &  ---  &  20.1 \\
   IRAS~60      & 60  &  --- &  ---  & 145   \\
   IRAS~100     & 100 &  --- &  ---  & 217   \\
   SCUBA 450    & 450 &   13 &  177  & 190   \\
   SCUBA 850    & 850 &   19 &   76  &  95   \\
                &     &      &       &       \\
   LW2/LW3      &     & 0.50 &  0.82 &  0.78 \\
   450/850      &     & 0.68 &  2.32 &  2.00 \\
  \hline
  \\
\end{tabular}
\noindent
\\
Note:\ $R<7.5$\arcsec\ (nucleus); $R=7.5$\arcsec - 60\arcsec\ (disk); 
$R<60$\arcsec\ (total)\\
\\
$^a${\ts}IRAS fluxes at 12\mic, 25\mic, 60\mic, and 100\mic\ were 
computed 
from processed maps using the maximum entropy deconvolution routine 
HIRES (\cite{Surace}), which provides a beamsize of $\sim$1.5$^\prime$.  It was 
impossible to give a reliable estimate for the IRAS flux within the 
central 15\arcsec\ region. 
\end{table}

\section{Discussion}

\subsection{The barred spiral}

\begin{figure*}
\resizebox{\hsize}{!}{\includegraphics{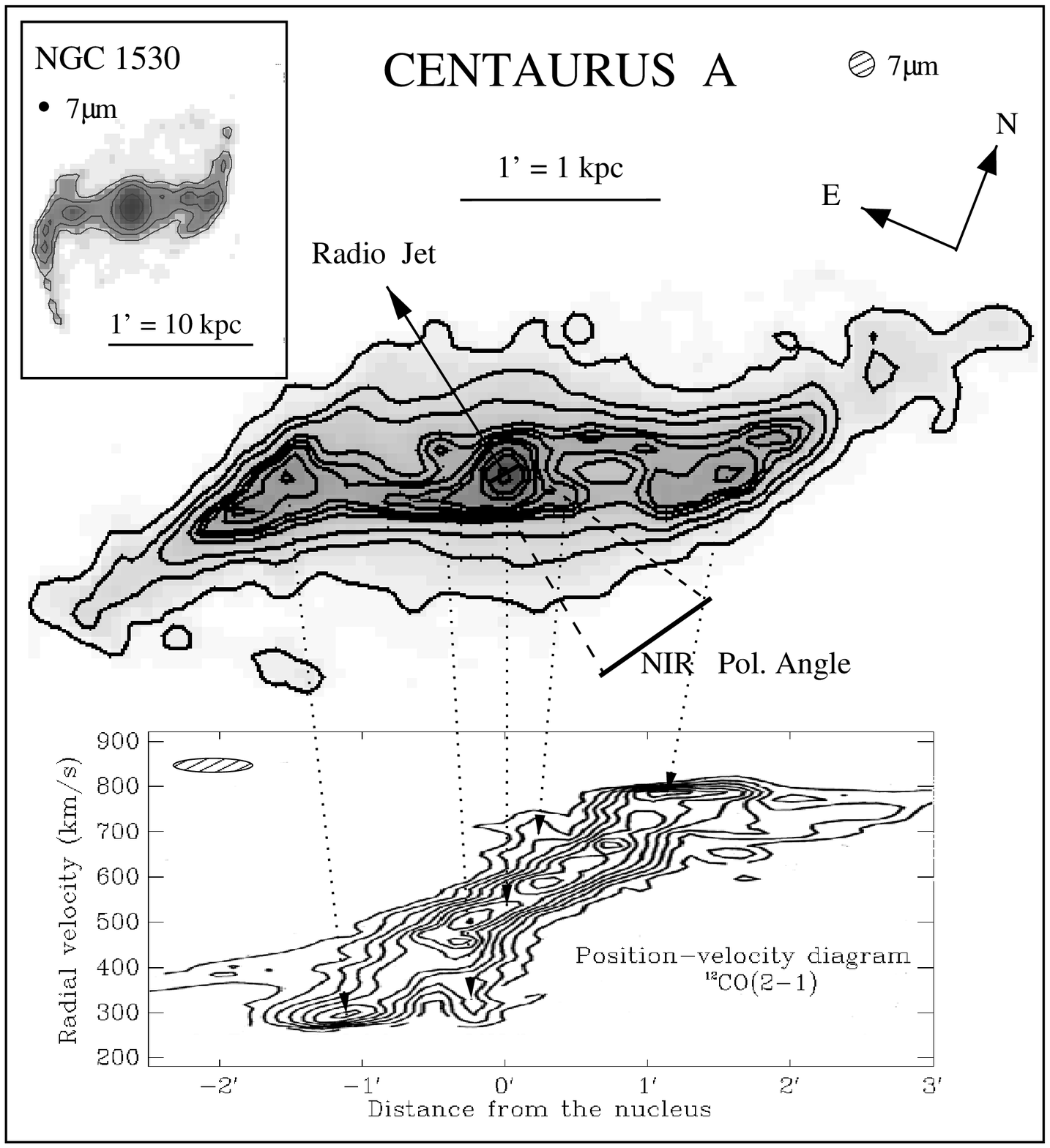}}
\caption{
ISO 7{\ts}$\mu$m image and $^{12}$CO(2-1) position-velocity map (along
PA -63.5$^{\circ}$ with FWHM=30\arcsec, \cite{Quillen1}) of the
central region of CenA.  Note the similarity in morphology with the
7{\ts}$\mu$m image of the prototype barred spiral NGC{\ts}1530
(courtesy of C. Bonoli) shown in the upper inset. In NGC{\ts}1530 the
plane that contains the bar is tilted by $\sim${\ts}55$^{\circ}$ to
the line of sight (\cite{Reynaud}), whereas in CenA it is tilted by
$\sim${\ts}18$^{\circ}$ (\cite{Dufour}, \cite{Graham}).  The overall
structure exhibited by the 7{\ts}$\mu$m emission from CenA is that of
a barred spiral with a primary bar extending $\sim${\ts}1$^\prime$ in
radius from the nucleus, connected in its outer ends to trailing
spiral arms. The strongest 7{\ts}$\mu$m emission from the primary bar
is along its leading edge in what takes the form of two slightly
curved arcs, where shocks and density enhancements take place. The
inner ends of these two arcs are connected to what may be a secondary
nuclear bar whose position angle is defined by the NIR
K(2.2{\ts}$\mu$m) band polarization (\cite{Packham}).  On the plane of
the sky the radio jets appear perpendicular to the innermost
polarization angle.  The kinematics of the gas in the lower panel is
consistent with a barred spiral, where the bar rotates as a rigid body
within 70$^{\prime\prime}$, whereas at radii larger than
$\pm${\ts}70$^{\prime\prime}$ the gas exhibits the differential
rotation (flat rotation curve) typical of galactic disks. An apparent
E--W high velocity feature inside a radius of
$\sim${\ts}20$^{\prime\prime}$ with a high velocity extent of
$\sim${\ts}280 to 750 km s$^{-1}$ in the $^{12}$CO(2-1) emission is
not accounted for in the above description.
}
\label{composite}
\end{figure*}

In the following (see Figure \ref{composite}) it is shown that the
interpretation of the bisymmetric structure, observed at the centre of
CenA as a barred spiral is fully consistent with: 1) the morphology of
the dust lanes, observed in galaxies classified as barred spirals
(such as NGC 1530), 2) the kinematics observed in CO data
(\cite{Eckart2}, \cite{Quillen1}) and 3) theoretical models
(\cite{Athanassoula}) that predict shocks at the leading edge of bars,
producing an arc-like appearance of the warm dust.

The overall structure exhibited by the 7{\ts}$\mu$m emission from CenA
is that of a barred spiral with a primary bar extending
$\sim${\ts}1$^\prime$ in radius from the nucleus, connected in its
outer ends to trailing spiral arms that have the typical structures
seen in barred spirals with primary bars that end near, but somewhat
inside, their corotation resonance. The strongest 7{\ts}$\mu$m
emission from the primary bar is along its leading edge in what takes
the form of two slightly curved arcs, where shocks and density
enhancements should take place according to theoretical models
(\cite{Athanassoula}). The southeast arc moves toward the observer and
is in the foreground side, whereas the northwest arc moves away and is
in the background. The inner ends of these two arcs are connected to
what may be a secondary nuclear bar whose position angle is defined by
the NIR K(2.2{\ts}$\mu$m) band polarization (\cite{Packham}).  On the
plane of the sky the radio jets appear perpendicular to the innermost
polarization angle. The rotation in NGC 1530 is clockwise whereas that
in CenA is anticlockwise, and the appearance of the trailing arms in
CenA can be understood if at the ends of the primary bar (r $\sim$ 70
\arcsec) there is an increase in the inclination due to a strong warp
in the disk.  This is suggested by the apparent drop in the radial
velocities beyond r $\sim$ 70 \arcsec (see
Fig. \ref{composite}). Because of this strong warp the winding
structure of the spiral arms in CenA has a leading appearance.  The
kinematics of the gas in the lower panel is consistent with a barred
spiral, where the bar rotates as a rigid body within
70$^{\prime\prime}$, whereas at radii larger than
$\pm${\ts}70$^{\prime\prime}$ the gas exhibits the differential
rotation (flat rotation curve) typical of galactic disks. 
An apparent E--W high velocity feature inside a radius of
$\sim${\ts}20$^{\prime\prime}$ with a high velocity extent of
$\sim${\ts}280 to 750 km s$^{-1}$in the $^{12}$CO(2-1) emission is not
accounted for in the above description.

ISOCAM can not resolve features smaller than $\sim$ 5\arcsec\ but the
vectors of polarization at K(2.2{\ts}$\mu$m) which are produced by
absorption of the radiation from stars embedded in the dust lane
(\cite{Packham}) can be used to trace with higher angular resolution
the preferential distribution of dust in the innermost central region.
The position angle of the polarization shown in Figure \ref{composite}
suggests the presence of a secondary bar, or ``nuclear'' bar, of gas
of few hundred parsecs in size. This presumed secondary bar inside the
primary bar could be the dynamical instability that brings gas towards
the supermassive black hole (\cite{Shlosman}).  In fact, molecular gas
absorption has been detected in front of the compact nuclear source at
millimeter wavelengths (\cite{Eckart1}, \cite{Israel},
\cite{Quillen1}, \cite{Wiklind}), and it has been 
proposed (\cite{Israel}) that the absorption at redshifted velocities 
represents gas falling into the centre.

The jets and the near infrared polarization angle of the innermost
region appear to be perpendicular on the plane of the sky (Figure
\ref{composite}). However, the NE jet lies in the foreground and when
deprojected from the plane of the sky, the jets may not necessarily be
perpendicular to the nuclear bar.  Hubble Space Telescope (HST)
observations of Pa$\alpha$ ($\lambda = 1.87${\ts}$\mu$m) have shown a
structure with an elongation of 2\arcsec\ that has been interpreted as
an inclined nuclear disk of ionized gas (\cite{Schreier}), but it is
not perpendicular to the radio jets.

The ISOCAM observations of CenA show more details in the inner
morphology of the interstellar matter than previous single-dish
millimeter observations of the molecular gas (\cite{Eckart2},
\cite{Israel}, \cite{Quillen1}, \cite{Rydbeck}). However, the more
general characteristics of the circumnuclear molecular structure that
were proposed from CO(2-1) observations with a beamwidth (FWHM) of
22\arcsec\ using deconvolution techniques (\cite{Rydbeck}), and from a
variety of molecular lines (\cite{Israel}), have general resemblances
with the structures now seen with higher angular resolution in the
mid-infrared.  We point out that the interpretation of a barred spiral
in the inner region of CenA is not in conflict, and is even fully
consistent with the presence of a warped disk at larger radii. The
barred spiral is a dynamic instability, i.e. a density wave in the
warped disk of gas and dust.

\subsection{The formation and survival of the barred spiral}

It is believed that the same accretion event(s) that began tearing
apart a gas-rich object(s) also created the faint stellar
(\cite{Malin}) and gaseous (\cite{Schiminovich}) shells observed
around CenA.  It is known that the accretion of a small disk galaxy by
a massive elliptical would lead to the complete tidal disruption of
the former as it spirals inward in the potential of the elliptical
galaxy (\cite{Hernquist}). In this process the gas decouples from the
stars and sinks more readily to the centre (\cite{Barnes}), forming a
new disk out of the gaseous component alone. In CenA the overall
angular momentum of the newly formed disk is not aligned with the
major axis of the elliptical.  Therefore, the gaseous disk is subject
to torques forcing it to warp (\cite{Barnes}). Although gas is still
settling towards the central regions, the morphological and dynamical
symmetry of the spiral indicate that it is a stable structure and not
a transient feature. Rotating at 250 km s$^{-1}$ it must have
undergone several full rotations depending on how long after the
initial encounter it took the gas to settle into the central disk we
now see.

Using near infrared photometry (\cite{Quillen2}), as well as the
kinematics of the gas, a disk-to-total mass ratio within the turnover
radius of the rotation curve (within a sphere of 70 \arcsec) of the
order of $10^{-2}$ is obtained. N-body simulations would rule out that
a low-mass stellar bar has formed spontaneously in the disk, and
survived at steady-state.  However these simulations consider stellar
disks, while here we are dealing with a gaseous one.  Assuming a much
smaller velocity dispersion for the molecular gas of 5-10 km s$^{-1}$
(compared with 50 km s$^{-1}$ for a stellar disk and 145 km s$^{-1}$
for the spheroidal component in CenA, \cite{Wilkinson},
\cite{Eckart1}), we derive a much lower Toomre's Q parameter (of the
order of 1) for the gaseous disk in CenA. Therefore, a gaseous disk is
much more self-gravitating than a stellar one of similar
mass. Consequently, the bar in CenA can be in a quasi-steady state and
might have been formed spontaneously, or be driven.

\subsection{The symbiotic galaxy CenA: A template for giant radio 
galaxies}

Probably CenA is not an exception and could serve as a well-positioned
template to examine in detail the clues to the origin and evolution of
activity in early-type radio galaxies with similar radio morphology,
namely, with giant double radio lobes. This class of radio galaxies
can only be produced if there is continuous injection of relativistic
jet energy for $\geq$ 10$^8$ years, which requires enormous reservoirs
of interstellar gas and dust. In fact, prominent dust bands are
frequently observed in the hosts of giant radio galaxies. Fornax A,
the second nearest radio galaxy of this class, exhibits the dusty
signs for the merger of gas-rich galaxies on an early type galaxy.
Cygnus A, the prototype radio galaxy with double morphology, is
crossed by prominent optically dark bands that contain $\sim$10$^8$
M$_{\odot}$ of dust (\cite{Robson}). On the other hand, it has been
shown that in dusty radio galaxies with double radio structure, the
dust is usually found perpendicular to the radio axis, which suggests 
a connection between the mechanism leading to these double radio morphologies 
and the rotation axis of the dust
(\cite{Kotanyi},\cite{Dokkum}).

The specific mechanism in rapidly rotating disks of gas and dust that
brings fuel to the central engine in radio loud AGNs has been
difficult to probe observationally for several reasons. First,
galaxies similar to CenA are at greater distances (for instance,
Fornax A is 5-10 times and Cygnus A is $\sim$ 70 times more distant
than CenA), and at those distances it is difficult to see the detailed
morphology of dust and gas on scales $\leq$ 100 pc.  Second, at
optical and near-infrared wavelengths the light from the old stellar
population with a giant ellipsoidal distribution overwhelms any
emission from dust and newly formed stars in the deep interior. It is
in the mid-infrared that the emission from very warm dust can be
better traced, and to this end we had to wait for the unprecedented
capabilities of ISOCAM. Third, the observation of the cold gas
distribution by means of millimeter observations of weak molecular
line emission on top of the strong continuum of powerful radio
galaxies is a difficult task.

The barred spiral at the centre of CenA has dimensions comparable to
that of the small Local Group galaxy Messier 33, and is much larger
than the small dusty and/or ionized spiral features of 100-200 pc
radius seen with the HST in elliptical galaxies (\cite{Ford}).  It
lies on a plane that is almost parallel to the minor axis of the giant
elliptical. Whereas the spiral rotates with maximum radial velocities
of $\sim${\ts}250 km s$^{-1}$, the ellipsoidal stellar component seems
to rotate slowly (maximum line-of-sight velocity is $\sim${\ts}40 km
s$^{-1}$) approximately perpendicular to the dust lane
(\cite{Wilkinson}). The genesis, morphology, and dynamics of the
spiral formed at the centre of CenA are determined by the
gravitational potential of the elliptical, much as a usual spiral with
its dark matter halo. On the other hand, the AGN that powers the radio
jets is fed by gas funneled to the center via the bar structure of the
spiral. The spatial co-existence and intimate association between
these two distinct and dissimilar systems suggest a ``symbiotic"
association.

\section{Conclusions}

 From the observations reported here we can conclude that:

1) A quasi-stable barred spiral with the size of a small galaxy was
formed at the centre of the giant elliptical CenA, out of the tidal
debris of gas-rich accreted object(s).

2) The spiral has an infrared luminosity per unit mass of interstellar
molecular gas typical of starburst galaxies.

3) The bar may be the dynamical instability that serves to steadily
feed the AGN with the amount of fuel that is required to power the
giant radio lobes.

4) The spiral and the elliptical are distinct systems in symbiotic
association.

5) The observations reported here open the question on whether giant 
radio galaxies are in general, like CenA symbiotic galaxies.

\begin{acknowledgements} 
We thank E. Athanassoula, A. Bosma, J. Lequeux, M. Prieto,
D. Golombek, F.  Macchetto \& F. Combes for their valuable comments
and aid on different aspects of this work. ISO is an ESA project with
participation of ISAS and NASA.  The JCMT is operated by the
observatories on behalf of the UK Particle Physics and Astronomy
Research Council, the Netherlands Organization for Scientific
Research, and the Canadian National Research Council.
\end{acknowledgements}


\vfill\eject

\end{document}